# Dispersion Engineering of Planar Sub-millimeter Wave Waveguides and Resonators with Low Radiation Loss

Furkan Sahbaz[1-4] and Simeon I. Bogdanov[1-4]

[1]Department of Electrical and Computer Engineering, and Nick Holonyak, Jr. Micro and Nanotechnology Laboratory, University of Illinois at Urbana-Champaign, Urbana, Illinois 60801, USA

[2]Holonyak Micro and Nanotechnology Lab, University of Illinois at Urbana-Champaign, Urbana, Illinois 61801, USA

[3]Illinois Quantum Information Science and Technology Center, University of Illinois Urbana-Champaign, Urbana, Illinois 61801, USA

[4]The Grainger College of Engineering, University of Illinois Urbana-Champaign, Urbana, Illinois 61801, USA

Mm-wave and THz superconducting circuits find numerous applications in areas ranging from quantum information and sensing to high-energy physics. Planar THz transmission lines and resonators are fabrication-friendly, compact, and scalable, and they can be efficiently interfaced with external signals and controls. However, planar circuits radiate strongly at high frequencies, which precludes their use in loss-sensitive applications. Here, we present the design and characterization of planar dispersion-engineered transmission lines that effectively suppress radiation leakage in desired mm-wave bands. We extend this concept to design planar resonators with extremely low radiation leakage, resulting in radiation $Q$-factors above $10^6$ at 553 GHz. Low-loss planar THz circuitry will impact many application domains, including broadband communications, quantum information, radio astronomy, and cosmology.

## I. INTRODUCTION

THz superconducting circuits are emerging as key components in quantum information and networking [1–3], high-energy physics [4], radio astronomy [5], ultrafast communications [6], and superconducting computing [7]. Hybrid integration of low-loss, high-frequency superconducting circuitry with microwave and/or optical devices can unlock new regimes of operation, from quantum-limited transduction to axion detection [2,8–10]. Planar THz devices are inherently compatible with standard micro- and nanofabrication methods and offer scalable, on-chip integration with minimal footprint [11]. In addition, they provide straightforward interfacing with microwave and photonic elements to harness strong interaction within a wide range of frequencies [12,13]. On-chip applications requiring a balance between bandwidth and propagation losses, such as radio astronomy and communications, typically require loss tangents below $10^{-3}$ [14,15]. Moreover, for highly loss-sensitive applications like quantum information processing or axion detection, loss tangents below $10^{-6}$ are targeted [16,17]. Unfortunately, internal loss mechanisms in planar THz devices scale up steeply at high frequencies, limiting the range of feasible applications.

Significant progress has been made in optimizing transmission line and coupler geometries, as well as in tuning material properties to reduce losses in mm-wave superconducting circuits [18–20]. Internal quality factors ($Q_i$) above $10^5$ have been reported in ground-shielded geometries on sapphire, operating around 100 GHz [18]. However, as the operating frequency increases, additional loss mechanisms – particularly, radiation loss – become dominant, reducing the measured $Q_i$ by orders of magnitude (Fig.1). Specifically, mode leakage arising from phase velocity mismatch between the transmission line and the substrate is an inherent attribute of planar circuits.

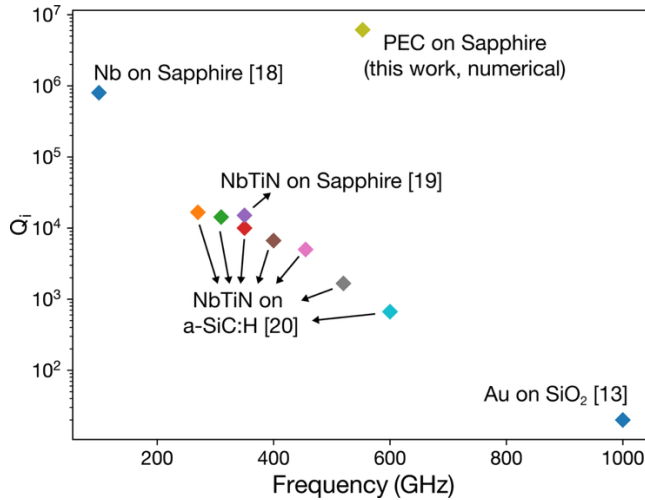

*Figure 1: Internal quality factors of superconducting waveguides and resonators calculated from the attenuation values reported in the literature [13,18-20].*

In this work, we show that by controlling the dispersion properties of transmission lines, one can dramatically suppress radiation and scattering losses in sub-THz waveguides and resonators. Dispersion engineering is a flexible and powerful tool employed in microwave and optical planar circuits to manipulate the frequency-dependent group and phase indices. Dispersion engineering achieves a range of useful effects such as phase shifts [21], delays [22], filtering parasitic signals [23], matching nonlinear interactions [24], and manipulating non-reciprocity [25]. Here, we use dispersion engineering, implemented through periodic and quasi-periodic modulations to strongly suppress radiation losses of CPS waveguides. By exploiting the rich parameter space of quasi-periodic waveguides, we show that it is possible to effectively control group velocity dispersion (GVD) and extend this method to design high-quality planar resonators using the so-called Gaussian mirrors [26]. As a result, radiation loss in CPS waveguides is reduced by more than two orders of magnitude, further enabling resonators with radiation quality factors over $10^7$.

## II. SCALING OF RADIATION LOSSES IN CPS WAVEGUIDES

As a starting point of our analysis, Eq. 1 [27] provides an intuitive guideline to understand the scaling of radiation loss ($\alpha_{rad}$) in a CPS waveguide as a function of frequency, mode index, and mode confinement. The radiation loss features a steep scaling with frequency ($\omega^3$ dependence), and with the mode size (dictated by $w$) analogously to the case of spontaneous emission.

$$\alpha_{rad} = \frac{\pi^2(3-\sqrt{8})}{16} \frac{n_p}{n_{sub}} \left(1 - \frac{n_p^2}{n_{sub}^2}\right)^2 \frac{(w+2w_e)n_{sub}^3}{c^3 K(k_e) K(\sqrt{1-k_e^2})} \omega^3 \quad (1)$$

At a given frequency, radiation losses are extremely sensitive to the phase index mismatch between the mode and the substrate ($n_p/n_{sub}$) as well as by the mode confinement described by the elliptic integral of the first kind $K(k_e)$, where $k_e = w/(w + 2w_e)$. To illustrate these effects, Figure 2 plots the frequency-dependent radiation losses (per mm) for four different waveguide configurations on sapphire substrate ($n_{sub}$ = 3.06). First, we consider waveguides with gap widths $w$ = 0.2 μm and $w$ = 2 μm (while keeping the total waveguide width fixed, $w + 2w_e$ = 6 μm). We also consider two cases of superstrates, air and SiO$_2$, where the latter effectively increases $n_p$ from 2.5 to 2.7.

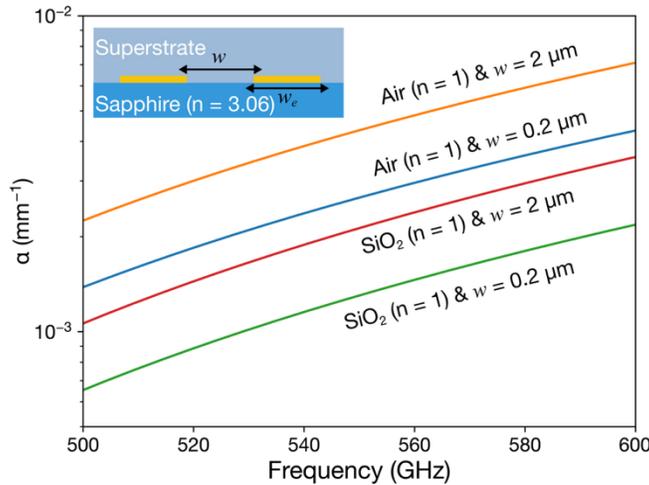

*Figure 2: Radiation loss for various $n_p$ values and waveguide dimensions. Inset is the cross section used to calculate radiation loss, while $w + 2w_e = 6$ μm is kept constant.*

A superstrate layer index-matched to the substrate can dramatically reduce the radiation losses. However, depending on the application, a closely index-matched superstrate may be impractical, especially when the CPS waveguide must be hybridized with external circuitry (e.g. microwave and/or optical). Therefore, here we focus on unmatched waveguides with a vacuum superstrate as the most unfavorable scenario. Mode confinement from $w$ = 2 μm to $w$ = 0.2 μm also somewhat reduces the radiation losses, but this effect comes at the expense of extensive nanofabrication along the entire length of the waveguide.

Periodic perturbations to the waveguide impedance can be used to modify its dispersion characteristics, typically through the creation of bandgaps. Specifically, close to the valence band edge, the dispersion deviates strongly from that of the uncorrugated waveguide, and both the phase index and the group index increase [28]. Crucially, even a small increase in the phase index can significantly suppress radiative loss, as suggested by Eq. 1. Modulating $w$ as a function of distance along the waveguide is a straightforward means to perturb the waveguide's impedance (see Fig. 3(a)). For example, a $w$ = 2 μm waveguide has a characteristic impedance of $Z_h$ = 100 Ω and a $w$ = 0.2 μm waveguide has $Z_l$ = 50 Ω. The dispersion properties of frequencies (denoted by their wavelength $\lambda$) in such waveguides can be estimated via

$$\cosh(k\Lambda) = \cos\left(\frac{2\pi n_h d_h}{\lambda}\right)\cos\left(\frac{2\pi n_l d_l}{\lambda}\right) - \frac{1}{2}\left(\frac{Z_h}{Z_l} + \frac{Z_l}{Z_h}\right)\sin\left(\frac{2\pi n_h d_h}{\lambda}\right)\sin\left(\frac{2\pi n_l d_l}{\lambda}\right) \quad (2)$$

where $k$ is the wavevector and $n_h$ ($n_l$) and $d_h$ ($d_l$) are the effective index and length corresponding to the waveguide cross-section with $Z_h$ ($Z_l$), respectively. By tuning the "groove length" $d_l$, periodicity $\Lambda$, and the contrast between impedances of waveguide cross sections ($Z_h/Z_l$), one can control the propagation characteristics near the valence band maximum (VBM). The crystal period $\Lambda$ dictates the mid-gap frequency. We examine the periodicity of $\Lambda$ = 100 μm corresponding to a mid-gap frequency of 657 GHz. We calculate the band structures for an infinitely long periodic structure using Eq. 2 for different values of $d_l$ (10 μm, 20 μm, 30 μm, and 40 μm) in Fig. 3(b). The resulting phase ($n_p$) and group ($n_g$) indices are plotted in Fig. 3(c)-(d). In addition, in Fig. 3(e), we plot the radiation loss in mm$^{-1}$ estimated using Eq. 1. Although strictly speaking Eq. 1 does not apply to structures with broken translation symmetry, it provides a valuable intuitive insight into the effect of the periodic perturbations that is confirmed by numerical simulations below.

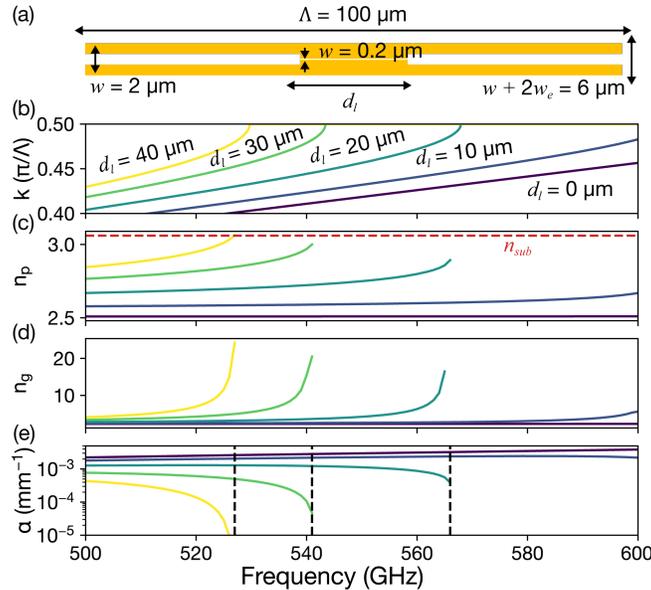

Figure 3: (a) A single unit cell in the infinitely long periodic structure used in calculations. (b) Calculated band structures for infinite structures with $d_l$ from 0 μm through 40 μm. Estimated (c) phase index, (d) group index, and (e)

*radiation loss for $d_l$ from 0 μm through 40 μm. The horizontal line in (c) represents the substrate phase index, whereas the vertical lines in (e) represent the valence band maxima of each band structure.*

With increasing $d_l$ values, the reflections at the grooves become stronger and the bandgap widens accordingly. The valence band departs further from the vacuum line, resulting in a higher phase index, approaching the value of $n_{sub}$. As an important caveat, at the band edge, the group index also diverges, and the waveguide operation becomes impractical due to diverging group velocity dispersion and material losses. However, the region with increasing phase index extends far below the VBM (indicated by the dashed vertical lines in Fig. 3(e)). Accordingly, the radiation losses calculated using Eq. 1, reduce by orders of magnitude in a frequency range on the order of 10 GHz, which constitutes a practical bandwidth for most sub-THz applications.

## III.   NUMERICAL SIMULATIONS OF FINITE WAVEGUIDES

Although Eq. 1 provides useful guidance for choosing the key waveguide design parameters, it lacks a complete description of radiation mechanisms. In particular, it must be modified to account for a lack of translational invariance and cannot properly describe signal reflection and out-of-plane scattering due to periodic corrugations. In order to provide a complete analysis of radiation in periodically modulated waveguides, we discuss the numerical simulations of periodic waveguides constructed from 12 periods of modulated gap with, as shown in Fig. 4, using COMSOL Multiphysics software [29]. To focus on radiation and scattering losses, we assume no material-based losses in this work and utilize perfect electric conductors, as well as lossless dielectrics, to model the devices. To suppress the scattering losses and resonant reflections, we apodize (smoothen) the corrugations as shown in Fig. 4(b) using 5 μm long impedance-matching transitions between the wide and narrow waveguide sections [28,30].

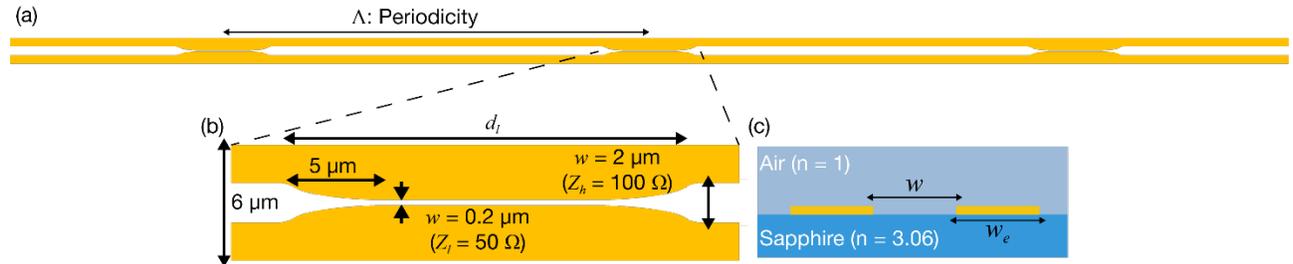

*Figure 4: (a) Top view, (b) a unit cell, and (c) cross section of the numerically simulated structures.*

We initially consider a fixed periodicity of Λ = 100 μm, to estimate radiation losses in line with previous calculations, yielding a total waveguide length of 1.2 mm along the *z*-axis. In the absence of material losses, radiation loss in the simulated setup can be extracted as the power that is not delivered to either of the ports as a reflected or transmitted signal, via $P_{rad} = P_{in}(1 - |S_{11}|^2 - |S_{21}|^2)$. See Appendix B for further details on further details on numerical calculations.

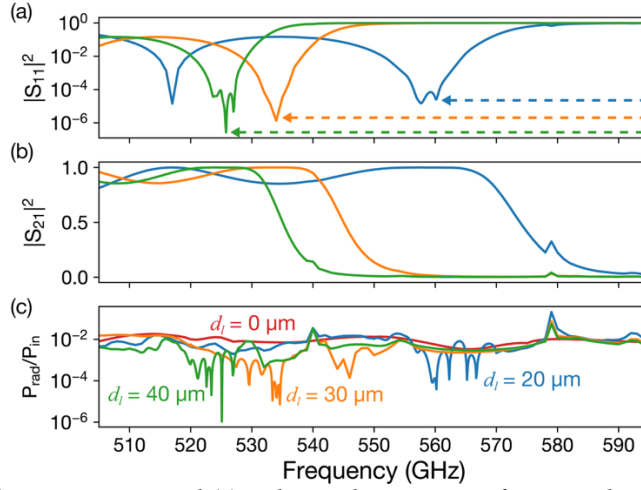

*Figure 5: (a) Reflection, (b) transmission, and (c) radiation loss spectra of waveguides that are constructed with 12 periods of Λ = 100 μm and $d_l$ = 20 μm, 30μm and 40 μm. Radiation loss of a bare waveguide (w = 2 μm) is plotted in (c) to serve as a reference. The dashed lines in (a) approximately indicate the valence band maxima.*

Figure 5(a)-(c) presents the numerically simulated $|S_{11}|^2$, $|S_{21}|^2$, and $P_{rad}$ spectra, respectively, for waveguides with 20 μm, 30 μm, and 40 μm-long grooves. Although the numerically simulated radiation spectra in Fig. 5(c) are slightly higher due to finite domain constraints and the presence of additional loss channels, they validate the qualitative insight given by Eq. 1. In particular, the radiation losses are strongly reduced in the vicinity of the VBM within a bandwidth of over 10 GHz. Moreover, local discontinuities at the grooves can lead to scattering and disrupt this reduction in radiated power. Periodic grooves without proper apodization (with sharp corners) strongly reduce the bandwidth of the loss suppression (Appendix A), compared to those with "smoothed" corners as in Fig. 4(b). It is therefore crucial to eliminate discontinuities along the periodic grooves in order to achieve low radiation loss [30].

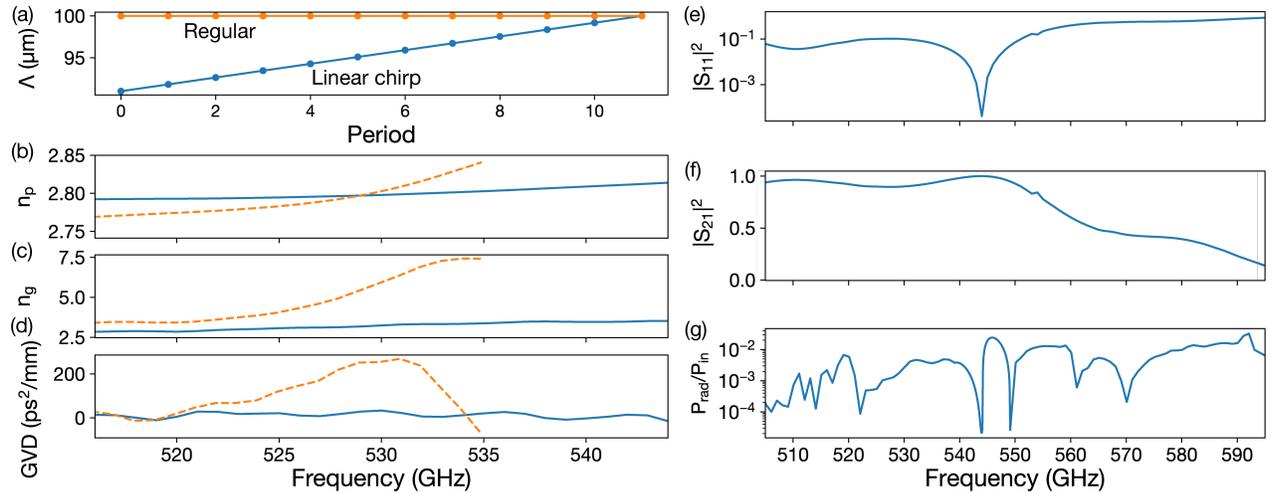

*Figure 6: (a) Periodicity profiles and the corresponding numerically calculated (b) phase index, (c) group index, and (d) group velocity dispersion for regular and chirped periodic waveguides. Similarly, (e) reflection, (f) transmission, and (g) radiation loss profiles in chirped periodic waveguide.*

An undesirable effect of bandgap proximity is a resonant increase in group index and group velocity dispersion (GVD) near VBM. Figures 6(b)-(d) show the changes in phase and group indices, as well as the resulting GVD at frequencies up to the VBM of periodic structures with $d_l$ = 30 µm. The accumulated phases were extracted from the numerically simulated transmission spectrum of each waveguide. Moreover, the GVD plotted in Fig. 6(d) (dashed lines) can lead to increased material losses and pulse distortion. Fortunately, a small departure from periodicity effectively mitigates these issues. By slowly perturbing the periodicity throughout the waveguide, one can design segments with different "local" band structures. In particular, a linearly graded periodicity strongly reduces the GVD due to the flattening of the phase index dispersion resulting from averaging over local band structures. Furthermore, using segments with positive GVD (valence band states) and negative GVD (where the operating frequency falls locally inside the bandgap), one can achieve dramatic GVD reduction through compensation. Reflections usually associated with frequencies inside the bandgap [31] are avoided because the varying periodicity along the propagation direction modifies the coupling, between forward- and backward-propagating modes [32–34]. As a result, for some frequencies that nominally fall inside the bandgap for a portion of the graded structure, the reflection coefficients are instead strongly canceled. Figure 6 provides a comparison between the periodic waveguides with regular periodicity ($\Lambda$ = 100 µm) and linear chirp ($\Lambda$ increased by 9 µm over 12 periods) for $d_l$ = 30 µm. The phase index, shown up to the modified VBM, follows an approximately linear profile near the band gap, leading to stabilization of the group index that is nearly constant in frequency. As a result, GVD remains below 10 ps$^2$/mm (and down to $10^{-1}$ ps$^2$/mm) within 10 GHz of the VBM. Even though the improvement in phase index is not as strong as in the case of a periodic structure, the achievable values can still effectively suppress radiation losses.

## IV. DESIGN AND NUMERICAL SIMULATION OF A WAVEGUIDE-COUPLED CAVITY

Quasi-periodically patterned waveguides harbor a rich set of tools to control the characteristics of propagating modes. In this section, we show that these structures can be used to construct cavities with internal (radiative) Q-factors exceeding $10^6$ at frequencies above 500 GHz. They can also be combined with waveguides for low-loss coupling and signal delivery. To analyze the contribution of radiation in cavity losses, we decompose the loaded Q-factor ($Q_l$) in terms of the external coupling ($Q_e$) on the one hand, and the material absorption ($Q_m$), radiation ($Q_{rad,wg}$) and mirror scattering ($Q_{rad,mir}$) constituting the internal Q-factor $Q_i$.

$$\frac{1}{Q_l} = \frac{1}{Q_m} + \frac{1}{Q_{rad,wg}} + \frac{1}{Q_{rad,mir}} + \frac{1}{Q_e} = \frac{1}{Q_i} + \frac{1}{Q_e} \quad (4)$$

As our model includes no material-based losses, the absorption term $1/Q_m$ arising from ohmic, dielectric, and two-level systems losses is neglected. Constructing a cavity by directly introducing a defect region inside periodic waveguides (e.g. by regular Bragg mirrors) typically leads to severe radiation caused by mirror scattering, reducing $Q_{rad,mir}$ [35]. Generally, due to the attenuation of the mode at the mirrors, some spatial Fourier components of the cavity mode end up in the light cone and can scatter out of plane. It has been shown in optical photonic crystal cavities that mirrors with Gaussian field attenuation profile can eliminate the Fourier components in the leaky region and therefore minimize the additional radiation loss due to mirrors [26,36].

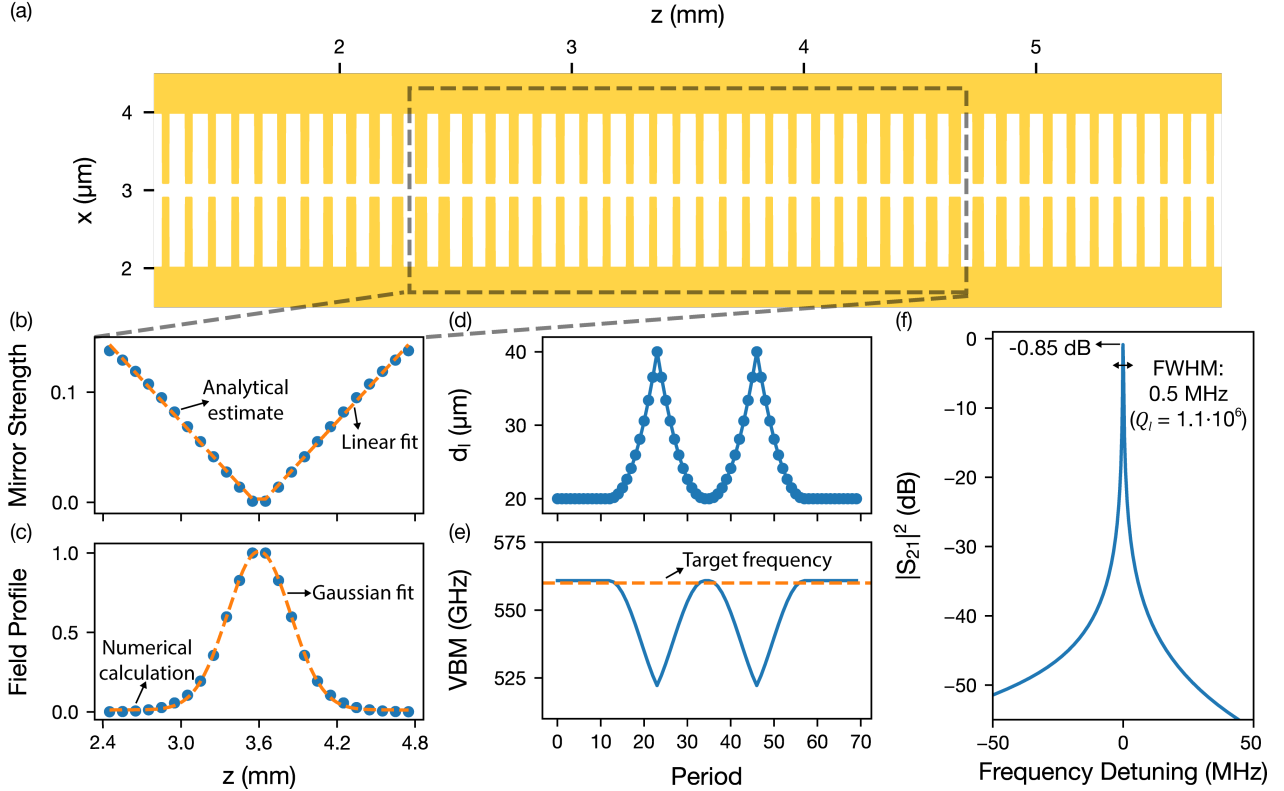

*Figure 7: (a) Cavity coupled to periodically modulated waveguides on each side (12 periods with Λ = 100 μm and $d_l$ = 20 μm, not shown). See Appendix B for the full structure. (b) Analytically estimated mirror strength and (c) numerical simulations of field attenuation profile in the cavity region (dashed area). Evolution of (d) $d_l$ and (e) corresponding valence band maxima throughout the structure. (f) Numerically simulated transmission spectrum of the waveguide-coupled cavity.*

We consider a photonic crystal cavity coupled to a periodic waveguide, as shown in Fig. 7 (Λ = 100 μm). Initially, the operating frequency is chosen slightly below the VBM of a periodically modulated waveguide with $d_l$ = 20 μm (∼ 560 GHz). The $d_l$ = 20 μm waveguide thus features a propagating mode at this frequency, corresponding to zero mirror strength. In contrast, a structure with $d_l$ = 40 μm features a high mirror strength of ∼ 0.13 at that frequency. A quadratic tapering of the modulation strength ($d_l$) from the outer end of the mirror towards the inner end leads to a linearly reducing mirror strength towards the cavity center, Fig. 7(b), and a Gaussian field profile of the cavity mode, Fig. 7(c). It has been previously shown that $L = 0$ cavities with no additional spacing between the mirror regions (with undisturbed periodicities) are the most effective in lowering radiation [36], so the $d_l$ = 20 um section is kept at a minimum length of a single period.

The resonant signal is coupled back evanescently into a periodically modulated waveguide. To avoid significant mode impedance mismatch and reflections at the waveguide-cavity interface, we employ the same tapering in $d_l$ (40 μm → 20 μm) while transitioning from the cavity to the periodic waveguide, as shown in Fig. 7(d)-(e). The gradual reduction in modulation strength ensures that the mirror strength is reduced linearly from the cavity to the waveguide, and that the resonant mode is transferred smoothly without introducing significant mismatch. Numerical simulations of the coupler region show around 90% coupling efficiency, confirming that our structure is sufficiently adiabatic.

Numerical simulations of the full cavity structure indicate a resonance frequency around 553 GHz (within ~1% of the numerically simulated band edge), and a full width at half maximum of approximately 0.5 MHz, leading to $Q_l$ of $1.1 \cdot 10^6$. The simulated transmission peak in Fig. 7(f) is approximately 0.82, confirming that the loss introduced by the couplers is negligible. $Q_e$, dictated by the mirror strength as well as coupling to the waveguide, can be estimated via $\left|S_{21}^{peak}\right| = Q_l/Q_e$ as $1.34 \cdot 10^6$, leading to $Q_i$ of approximately $6.14 \cdot 10^6$ ($Q_i > Q_e$). As a result, we conclude that planar resonators constructed with (periodic) Gaussian mirrors promise to show over three orders of magnitude improvement in $Q_i$, compared to planar Fabry-Perot resonators operating at similar frequencies [19,20].

## V. DISCUSSION

Planar superconducting circuitry with ultra-low radiative loss characteristics can open extensive new application possibilities in the sub-THz range, including quantum information processing, radio astronomy, and high-energy physics. Specifically, the waveguide and cavity designs outlined in this paper can be useful in quantum information processing to implement high-frequency qubits [3] or transducers [2]. In high-energy physics and radio astronomy, where extremely sensitive detectors in the THz range are desirable, the dispersion-engineered elements can help extend the axion search to the meV range [4,5,37] with quantum-limited sensitivities [8].

Despite their limitations due to GVD, periodic structures can still maintain GHz-range bandwidth and serve as circuit elements in on-chip THz spectroscopy [38] or ultrafast communication systems [39]. Even though mitigating the dispersion effects via linear chirp limits the usable bandwidth of modes with lower radiation losses, optimized tapers can be used to minimize reflection and loss at these frequencies. In addition, periodic structures with weaker perturbations (shorter $d_l$) can be combined with high kinetic inductance superconductors to avoid high GVD while improving the phase index. Combination of these mitigation techniques can extend the usable bandwidth of frequency modes near the VBM, that maintain low radiation losses.

The structures presented in this paper focused exclusively on 2 μm- and 0.2 μm-wide gaps and a total waveguide width of 6 μm. Such length-scales are compatible with standard planar nanofabrication tools and can readily be fabricated with high accuracy using current processing capabilities. However, the reduced radiation losses can also be achieved in different geometries. The dispersion relation outlined in Eq. 2 indicates that, as long as similar contrast between impedances or the effective indices is maintained, waveguides with different aspect ratios (e.g. wider gaps, shorter grooves) can be tailored to maintain similar properties.

**ACKNOWLEDGEMENTS**

We acknowledge William J. Cabal II for help in preparing the manuscript. F.S. and S.I.B. acknowledge the startup funding at the University of Illinois at Urbana-Champaign.

**DATA AVAILABILITY**

The data that support the findings of this article are openly available [40].

# APPENDIX A: WAVEGUIDES WITHOUT GROOVE APODIZATION

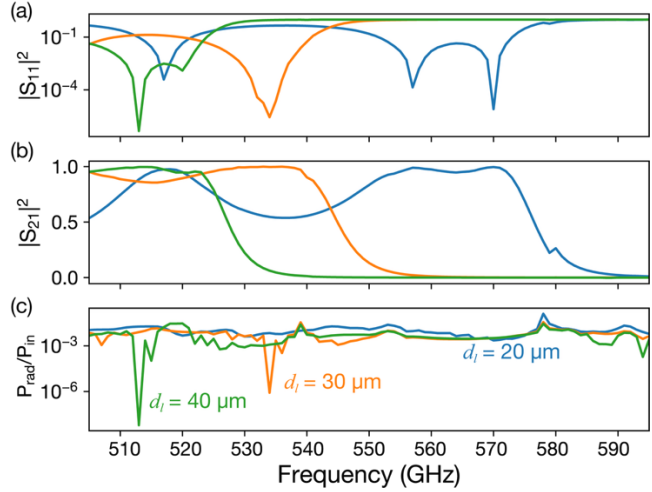

*Figure 8: (a) Reflection, (b) transmission, and (c) radiation loss spectra of waveguides without groove apodization.*

# APPENDIX B: CAVITY CONSTRUCTION VIA ABCD MATRICES

In order for an accurate measurement of radiation losses, the final simulation domain was formed as a "box" of dimensions 1.2 mm x 1 mm x 1.2 mm, surrounded by scattering boundary conditions except for the port boundaries at both ends of the waveguide. In order to simplify the calculations and use finer meshing, the domain was cut in half along the propagation direction (z). Furthermore, complex structure simulations that require significant computational resources (e.g. the waveguide-coupled cavity) were split into individual parts, e.g. waveguide, coupler, and cavity, then combined using ABCD matrices [28].

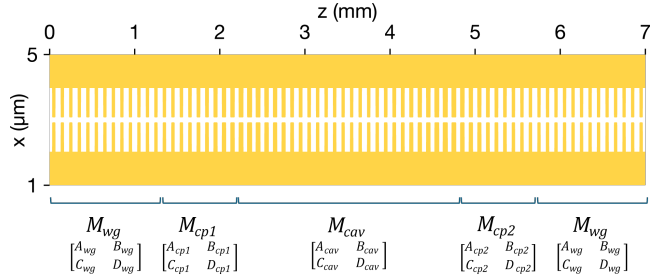

*Figure 9: Individual matrix components of the coupled cavity structure.*

Figure 9 shows the individual matrix elements of the waveguide-coupled cavity presented in Fig. 7, where $M_{wg}$, $M_{cav}$, and $M_{cpi}$ ($i \in \{1,2\}$) are the ABCD matrices corresponding to the waveguide, cavity, and couplers, respectively. The resulting structure can be represented by $M'$:

$$M' = M_{wg} M_{cp1} M_{cav} M_{cp2} M_{wg} = \begin{bmatrix} A' & B' \\ C' & D' \end{bmatrix} \quad (C1)$$

Finally, the $S_{11}$ and $S_{21}$ spectra can be estimated using the elements of $M'$, where $Z_0$ is the characteristic impedance [28]:

$$S_{11} = \frac{A' + \dfrac{B'}{Z_0} - C'Z_0 - D}{A' + \dfrac{B'}{Z_0} + C'Z_0 + D} \quad (C2.1)$$

$$S_{21} = \frac{2(A'D' - B'C')}{A' + \dfrac{B'}{Z_0} + C'Z_0 + D} \quad (C2.2)$$